%% file: ms.tex
\documentclass[iop]{emulateapj}
\usepackage{amsmath}

\input{newcom}

\newcommand{\Tabs}{\ensuremath{T_\mathrm{abs}}}
\newcommand{\Temis}{\ensuremath{T_\mathrm{emis}}}
\newcommand{\posterior}{\ensuremath{P(\gamma|\boldsymbol{d},\boldsymbol{m},\mathcal{I})}}
\newcommand{\likelihood}{\ensuremath{P(\boldsymbol{d}|\gamma,\boldsymbol{m},\mathcal{I})}}
\newcommand{\prior}{\ensuremath{P(\gamma|\boldsymbol{m},\mathcal{I})}}
\newcommand{\chisqmin}{\ensuremath{\chi^2_\mathrm{min}}}

\shorttitle{HALO CONVECTIVE INSTABILITY}
\shortauthors{HENLEY \& SHELTON}

\begin{document}

\title{Is the Milky Way's Hot Halo Convectively Unstable?}
\author{David B. Henley and Robin L. Shelton}
\affil{Department of Physics and Astronomy, University of Georgia, Athens, GA 30602; dbh@physast.uga.edu}

\begin{abstract}
We investigate the convective stability of two popular types of model of the gas distribution in the
hot Galactic halo.
We first consider models in which the halo density and temperature decrease exponentially with
height above the disk. These halo models were created to account for the fact that, on some sight
lines, the halo's X-ray emission lines and absorption lines yield different temperatures, implying
that the halo is non-isothermal.
We show that the hot gas in these exponential models is convectively unstable if $\gamma<3/2$, where
$\gamma$ is the ratio of the temperature and density scale heights.
Using published measurements of $\gamma$ and its uncertainty, we use Bayes' Theorem to infer
posterior probability distributions for $\gamma$, and hence the probability that the halo is
convectively unstable for different sight lines. We find that, if these exponential models are good
descriptions of the hot halo gas, at least in the first few kiloparsecs from the plane, the hot halo
is reasonably likely to be convectively unstable on two of the three sight lines for which scale
height information is available.
We also consider more extended models of the halo. While isothermal halo models are convectively
stable if the density decreases with distance from the Galaxy, a model of an extended adiabatic halo
in hydrostatic equilibrium with the Galaxy's dark matter is on the boundary between stability and
instability. However, we find that radiative cooling may perturb this model in the direction of
convective instability.
If the Galactic halo is indeed convectively unstable, this would argue in favor of supernova
activity in the Galactic disk contributing to the heating the hot halo gas.
\end{abstract}


\section{INTRODUCTION}
\label{sec:Introduction}

X-ray observations indicate the presence of hot gas above the disk of our Galaxy, in the halo, with
temperatures of $\sim$$(\mbox{1--3})\times10^6$~K. Evidence for this hot gas comes from observations
of the diffuse soft X-ray background (SXRB) emission \citep[e.g.,][]{kuntz00,yoshino09,henley13a},
and absorption lines from highly ionized metals in the X-ray spectra of active galactic nuclei
\citep[e.g.,][]{nicastro02,rasmussen03,yao07a,hagihara10,gupta12,miller13}. The hot halo gas is
thought to be due to supernova-driven outflows from the Galactic disk
\citep[e.g.,][]{shapiro76,norman89,joung06,hill12} and/or infall of extragalactic gas
\citep[e.g.,][]{toft02,rasmussen09,crain10}, but the relative importance of these two processes
remains uncertain \citep{henley10b,henley13a}.

In this paper, we consider the question of energy transport in the hot halo. In particular, we
investigate whether or not the hot halo gas is convectively unstable.  In a so-called galactic
fountain \citep{shapiro76,bregman80}, a convective flow is expected to be set up by supernova
heating: the heated gas moves from the disk into the halo, either by superbubbles breaking out of
the disk \citep[e.g.,][]{maclow89} or by regions of hot gas rising buoyantly \citep{avillez01}, and
then subsequently cools and falls back to the disk. However, while several different models for the
density and temperature distributions of the hot halo have recently been proposed, to the best
of our knowledge no one has investigated the convective stability of these models. Our goal here is
to help build a more complete picture of the physical processes occurring in the halo.

Whether or not a halo model is convectively unstable depends on the density and temperature
distributions of the gas, and there is some disagreement regarding the best model to describe these
distributions.  Such models fall mainly into two broad categories, which we refer to as exponential
and extended halo models (in this paper, we will examine halo models from both
categories). \citet[and references therein]{wang12} argue in favor of the former type of model,
saying that the X-ray observations are best described by a model in which the hot halo plasma is
concentrated relatively close to the disk, with an exponential scale height of a few
kiloparsecs. However, others argue that the X-ray observations indicate a more extended hot halo
\citep[$\ga$100~\kpc\ in extent; e.g.,][]{gupta12,gupta13,miller13}. There is also indirect evidence
for an extended halo, such as the lack of gas in satellite galaxies and the confinement of
high-velocity clouds \citep[and references therein]{fang13}. In reality, the halo may consist of a
combination of exponentially distributed gas close to the disk and more extended, lower-density gas
\citep{yao07a}.

In Section~\ref{sec:Models}, we will discuss the arguments for and against the exponential and
extended halo models. As the global morphology and extent of the hot halo are uncertain, we will
consider the convective stability of both types of model, in Sections~\ref{sec:Exponential} and
\ref{sec:Extended}, respectively.  We discuss our results in Section~\ref{sec:Discussion}, and
conclude with a summary in Section~\ref{sec:Summary}.

\section{MODELS OF THE MILKY WAY'S HOT HALO}
\label{sec:Models}

\subsection{Exponential Halo Models}
\label{subsec:ExponentialModel}

\citet{yao07a} argued that the observed halo \OVII/\OVIII\ column density ratio (measured from the
high-resolution \chandra\ grating spectrum of Mrk~421) and the diffuse \OVII/\OVIII\ emission ratio
(from a microcalorimeter spectrum of the SXRB; \citealt{mccammon02}) implied different halo
temperatures, implying that the halo is non-isothermal (see the first row of
Table~\ref{tab:Obs}\footnote{Note that \citet{yao07a} did not take into account the foreground
  emission from the Local Bubble and/or from solar wind charge exchange
  \citep[e.g.,][]{smith07a,henley08a,gupta09b,koutroumpa11} when inferring the halo temperature from
  the \citet{mccammon02} emission spectrum. However, the temperature that they obtained has been
  confirmed by \citet{gupta13}, using \suzaku\ observations of the SXRB from fields close to the
  Mrk~421 sight line, and taking into account the foreground emission (see the second row of
  Table~\ref{tab:Obs}).}). Instead, \citet{yao07a} found that the observations could be described by
a disk-like halo model in which the number density, $n$, and temperature, $T$, decreased
exponentially with height, $z$:
\begin{eqnarray}
  n(z) &=& n_0 \exp \left( - \frac{z}{h_n \xi} \right),
  \label{eq:Density} \\
  T(z) &=& T_0 \exp \left( - \frac{z}{h_T \xi} \right),
  \label{eq:Temperature}
\end{eqnarray}
where $n_0$ and $T_0$ are the midplane values, $h_n$ and $h_T$ are the scale heights (assumed to be
positive), and $\xi$ is the hot gas filling factor (assumed to be a constant).

\begin{deluxetable*}{lccccc}
\tablewidth{0pt}
\tablecaption{Observed Halo Temperatures, Observed Values of $\gamma \equiv h_T/h_n$, and Convective Instability Probabilities\label{tab:Obs}}
\tablehead{
\colhead{Sight line}    & \colhead{$\log \Temis$}      & \colhead{$\log \Tabs$}      & \colhead{$\gamma$}   & \colhead{Reference}   & \colhead{$P(\gamma<3/2)$} \\
                        & \colhead{(K)}                & \colhead{(K)} \\
\colhead{(1)}           & \colhead{(2)}                & \colhead{(3)}               & \colhead{(4)}        & \colhead{(5)}         & \colhead{(6)}
}
\startdata
Mrk~421                 & $\sim$6.33                   & 6.16 (6.10, 6.21)           & 0.29 ($<1.63$)       & 1                     & 0.89 \\
                        & 6.32 (6.29, 6.35)            & 6.16 (6.08, 6.24)           & \nodata              & 2                     & \nodata \\
LMC~X-3                 & 6.38 (6.34, 6.40)            & 6.11 (5.90, 6.30)           & 0.5 (0.1, 1.7)       & 3                     & 0.87 \\
PKS~2155$-$304          & 6.33 (6.31, 6.35)            & 6.27 (6.24, 6.29)           & 2.44 (1.03, 3.55)    & 4                     & 0.15 \\
                        & 6.36 (6.34, 6.38)            & 6.27 (6.22, 6.32)           & \nodata              & 2                     & \nodata
\enddata
\tablerefs{1.~Temperatures from \citet{yao07a}; $\gamma$ from \citet{sakai12}.
  2.~\citet{gupta13}.
  3.~\citet{yao09}, Table~3. \Temis, \Tabs, and $\gamma$ are from rows~1--3 of that table, respectively.
  4.~\citet{hagihara10}. \Temis, \Tabs, and $\gamma$ are from row~3 of Table~4, row~2 of Table~3, and row~1 of Table~7, respectively.}
\tablecomments{Columns~2 and 3 contain the halo temperatures inferred from the emission spectrum and
  from the absorption lines, respectively, assuming an isothermal halo. Column 4 contains the ratio
  of the temperature and density scale heights. The values in parentheses are the 90\%\ confidence
  intervals ($1\sigma$ confidence intervals for ref.~2).
  Column 6 contains the probability that the hot halo is convectively unstable on a
  given sight line, under the assumption that the exponential halo models are accurate descriptions
  of the hot halo (see Section~\ref{subsec:Probability}).}
\end{deluxetable*}

Given a set of model parameters, this exponential halo model predicts an X-ray emission spectrum and
a set of ion column densities. Essentially, the X-ray emission spectrum and ion column densities are
obtained by integrating $n^2\varepsilon(T,E)$ and $n f_i(T)$, respectively, along the line of sight,
where $\varepsilon(T,E)$ is the emissivity at photon energy $E$, $f_i(T)$ is the ion fraction for
the ion in question, and $n$ and $T$ are obtained from Equations~(\ref{eq:Density}) and
(\ref{eq:Temperature}), respectively. By fitting these model predictions simultaneously to the X-ray
absorption and emission line data, one can constrain the midplane density and temperature and the
corresponding scale heights. \cite{yao07a} obtained scale heights of $\sim$1--2~\kpc, implying that
the hot halo gas is confined close to the disk.

The \citet{yao07a} model has subsequently been applied to the LMC~X-3 and PKS~2155$-$304 sight lines
\citep{yao09,hagihara10}, as well as to a re-analysis of the Mrk~421 sight line \citep{sakai12}. In
each of these subsequent analyses, the absorption line data were obtained from \chandra\ grating
spectra of the targets in question, while the emission data were obtained from nearby blank-sky
\suzaku\ fields. Unlike \citet{yao07a}, \citet{yao09} and \citet{hagihara10} accounted for the
foreground emission from the Local Bubble and/or from solar wind charge exchange when modeling the
X-ray emission spectra (\citealt{sakai12} did not mention the foreground emission in their brief
report of their analysis). Similarly to \citet{yao07a}, these studies typically found $\Tabs <
\Temis$ (see Table~\ref{tab:Obs}) and obtained halo scale heights of a few kiloparsecs. However, for
the PKS~2155$-$304 sight line, the difference between $\log \Tabs$ and $\log \Temis$ is smaller than
on the other sight lines. As a result, $h_T$ is somewhat larger on this sight line than on the other
sight lines ($\sim$6~\kpc, versus $\la$3~\kpc). This suggests that, on this sight line, the halo is
closer to being isothermal than on the other sight lines.

\subsection{Extended Halo Models}

While the above-described studies have argued in favor of a hot halo with a scale height of a few
\kpc, others have argued for a much more extended halo, as is expected from models of disk galaxy
formation \citep[e.g.,][]{crain10}. As noted in the Introduction, indirect evidence for an extended
halo comes from the lack of gas in satellite galaxies and from the confinement of high velocity
clouds \citep[and references therein]{fang13}.

\citet{gupta12} have argued that there is direct X-ray observational evidence for an extended hot
halo. They combined their average measurement of the halo \OVII\ column density (from
\chandra\ grating spectra) with the mean halo emission measure taken from the literature to infer
the extent, $L$, of the hot halo gas. For an isothermal plasma of electron density \Ne, the column
density of a given ion $N \propto \Ne L$, while the emission measure $\mathcal{E} = \Ne^2 L$, and so
$L \propto N^2 / \mathcal{E}$. \citet{gupta12} found $L > 100~\kpc$, in stark contrast to the
results described above. A follow-up study, in which the emission measures were obtained from
blank-sky \suzaku\ fields adjacent to the sight lines from which column densities were obtained,
also found best-fit values of $L$ exceeding 100~\kpc\ \citep{gupta13}. \citet{gupta13} compared
their calculation of $L$ for the PKS 2155$-$304 sight line with that of \citet{hagihara10}
($109^{+200}_{-80}$ versus $4.0^{+1.9}_{-1.4}$~\kpc), finding that the differences were due to
differences in the values of the measured \OVII\ column density, the assumed oxygen abundance, and
the assumed ionization fraction used in the calculations.

\citet{wang12} have disputed \citepossessive{gupta12} conclusions. They argued that it is
inappropriate to adopt the mean column density and emission measure to infer $L$, as there is
considerable sight line-to-sight line variation in the observed column densities and emission
measures. (This issue was overcome in the above-mentioned follow-up study; \citealt{gupta13}.)
\citet{wang12} also argued that it is also inappropriate to assume that the halo is isothermal,
since typically $\Tabs < \Temis$ (see Table~\ref{tab:Obs}). This latter point is important because
the \OVII\ line tends to sample lower temperatures in absorption than in emission (see Figure~2 in
\citealt{wang12}). If one assumes an isothermal halo, then given an \OVII\ column density and an
\OVII\ line intensity, the inferred value of $L$ will change by an order of magnitude if the assumed
temperature changes by just 0.2~dex. Note that \citet{gupta13} also assumed an isothermal halo,
although their values of \Tabs\ are 0.16 and 0.09~dex smaller than their values of \Temis\ for the
Mrk~421 and PKS~2155$-$304 sight lines, respectively, and the confidence intervals for \Tabs\ and
\Temis\ on each sight line do not overlap (see the second and fifth rows of our
Table~\ref{tab:Obs}).

\citet{miller13} used a set of \xmm\ \OVII\ equivalent width measurements to constrain the
geometry of the halo. They initially assumed a spherically symmetric $\beta$-model in which the halo
density, $n$, as a function of Galactocentric distance, $r$, is given by
\begin{equation}
  n(r) = n_0 \left[ 1 + \left( \frac{r}{r_c} \right)^2 \right]^{-3\beta/2},
  \label{eq:BetaModel}
\end{equation}
where the core density, $n_0$, the core radius, $r_c$, and the slope at large radii, $\beta$, were
free parameters. (They also examined a flattened variant of this model.)  They initially assumed
that the halo had a maximum radius of 200~\kpc. Having determined their best-fit parameters, they
adjusted the halo size until the fit became unacceptable (in terms of $\chi^2$), and in this way
concluded that the halo is at least 18~\kpc\ (99\%\ confidence level) in size. However, before their
fitting, \citet{miller13} added an additional uncertainty of $\approx$7~\mA\ to their equivalent
widths' measurement uncertainties, to account for intrinsic scatter due to substructure in the
halo. It is unclear how their constraint on the halo's extent would depend on the size of this
additional uncertainty.

\subsection{Summary of Models}
\label{subsec:ModelSummary}

In summary, the global morphology and extent of the Milky Way's hot halo remain uncertain.  Studies
just of the halo X-ray emission do not help clarify this issue. \citet{yoshino09} found that the
halo emission measure tends to decrease toward the Galactic poles, although the decrease is steeper
than that expected from a plane-parallel disk model. In \citet{henley13a}, on the other hand, we
found no such tendency for the emission measure to decrease toward the Galactic poles in our much
larger data set, contrary to what is expected from a plane-parallel disk model. However, we pointed
out that the patchiness of the halo emission made it difficult to determine the halo's global
morphology. In reality, the hot halo may consist of higher-metallicity gas with a scale height of a
few kiloparsecs, which would tend to dominate the X-ray observations, plus a much lower-density,
lower-metallicity halo extending to $\sim$100~kpc, as inferred from indirect evidence
\citep{yao07a}.

As noted in the Introduction, in this paper we will first consider the exponential halo models
introduced by \citet{yao07a}, and examine the implications of the published results based on these
models (Section~\ref{sec:Exponential}). We will then look at more extended halo models in
Section~\ref{sec:Extended}.

\section{EXPONENTIAL HOT HALO MODEL}
\label{sec:Exponential}

\subsection{Convective Instability Criterion}
\label{subsec:Instability}

Here, we consider the convective stability of the exponential halo models described by
Equations~(\ref{eq:Density}) and (\ref{eq:Temperature}). In general, a gas will be convectively
unstable if the specific entropy, $\mathcal{S}$, decreases with height
\citep[e.g.,][Chapter~8]{shu92}. To simplify the mathematics, we define the following quantity:
\begin{equation}
  s = T n^{-2/3},
  \label{eq:Entropy}
\end{equation}
which is a monotonic function of $\mathcal{S}$ ($s=\alpha\exp(\mathcal{S}/\beta)$, where $\alpha$
and $\beta$ are constants). Since $s$ is a monotonic function of $\mathcal{S}$, the above convective
instability criterion applies to $s$, as well as to $\mathcal{S}$. For most of the remainder of this
paper, we will refer to $s$ as the entropy (though we will return to considering $\mathcal{S}$ in
Section~\ref{sec:Extended}).

Substituting Equations~(\ref{eq:Density}) and (\ref{eq:Temperature}) into Equation~(\ref{eq:Entropy}),
we obtain the entropy of \citepossessive{yao07a} exponential halo model as a function of height:
\begin{equation}
  s(z) = s_0 \exp \left( - \frac{z}{h_s \xi} \right),
  \label{eq:Entropy2}
\end{equation}
where
\begin{equation}
  s_0 = T_0 n_0^{-2/3}
\end{equation}
is the midplane entropy and
\begin{equation}
  h_s = \frac{3 h_T}{3 - 2 h_T/h_n}
  \label{eq:EntropyScaleHeight}
\end{equation}
is the entropy scale height.

If $h_s>0$, the entropy decreases with height (Equation~(\ref{eq:Entropy2})), and hence the hot halo
gas is convectively unstable. From Equation~(\ref{eq:EntropyScaleHeight}), we see that this occurs
if
\begin{equation}
  \gamma \equiv \frac{h_T}{h_n} < \frac{3}{2}.
  \label{eq:Instability}
\end{equation}
Thus, we have derived the convective instability criterion for \citepossessive{yao07a} exponential
halo model, in terms of the ratio of the temperature and density scale heights, $\gamma$. Note that
this criterion is independent of the filling factor, $\xi$, which is not constrained in the observational
analyses.

\subsection{Observations of $\gamma$}
\label{subsec:Observations}

We will use the results obtained by applying the \citet{yao07a} exponential halo model to joint
analyses of X-ray absorption lines in high-resolution \chandra\ grating spectra and of X-ray
emission from nearby \suzaku\ blank-sky fields \citep{yao09,hagihara10,sakai12}.\footnote{See
  Section~\ref{sec:Models} for a brief description of how these analyses constrained the exponential
  model's parameters.} As noted in Section~\ref{subsec:ExponentialModel}, \citet{yao09} and
\citet{hagihara10} accounted for the foreground X-ray emission in their analyses, so the best-fit
parameters for the exponential model are applicable to the halo. (As also noted in
Section~\ref{subsec:ExponentialModel}, \citet{sakai12} did not mention the foreground emission in
their report.)

The above studies obtained midplane densities of $\sim$$1\times10^{-3}~\mathrm{cm}^{-3}$, midplane
temperatures of $\sim$$3\times10^6~\mathrm{K}$, and scale heights of a few kiloparsecs. Here,
however, we are particularly interested in the ratio of the temperature and density scale heights,
$\gamma$. The measurements of $\gamma$ from these studies are shown in column~4 of
Table~\ref{tab:Obs}.

The low values of $\gamma$ measured toward Mrk~421 and LMC~X-3 seem to favor the hot halo being
convectively unstable (Equation~(\ref{eq:Instability})). However, because of the large upper error
bars, the $\gamma$s are also consistent with convective stability. On the PKS~2155$-$304 sight line,
meanwhile, the measured $\gamma$ favors convective stability, but is also consistent with convective
instability.

While the published values of $\gamma$ do not indicate unambiguously whether or not the hot halo is
convectively unstable, we can use the published uncertainties on $\gamma$ in combination with Bayes'
Theorem \citep[e.g.,][]{sivia06} to estimate the probability that the halo meets the criterion for
convective instability. We do this in the following section.

\subsection{Probabilities of Convective Instability}
\label{subsec:Probability}

To estimate the probability that the halo is convectively unstable on a given sight line, we first
estimate the posterior probability distribution for $\gamma$, \posterior. Here, $\boldsymbol{d}$
represents the observational data, $\boldsymbol{m}$ represents the other model parameters besides
$\gamma$, and $\mathcal{I}$ represents prior information and assumptions. Note that, for our
purposes, $\mathcal{I}$ includes the assumption that the halo exponential models are accurate
descriptions of the hot halo. Once we have estimated \posterior, we can estimate the probability
that $\gamma < 3/2$, and hence the probability of the proposition that the hot halo is convectively
unstable.

Bayes' Theorem is
\begin{equation}
  \posterior \propto \likelihood \prior,
  \label{eq:Bayes}
\end{equation}
where \likelihood\ and \prior\ are the likelihood and the prior probability, respectively. The likelihood
is the probability of obtaining the observed data, given the assumed model. The prior represents
our knowledge of $\gamma$ before the observations were taken. Other than requiring $\gamma$ to
be positive, we assume no prior knowledge of $\gamma$, and hence a uniform prior probability
distribution:
\begin{equation}
  \prior = \left\{
  \begin{array}{ll}
    \mbox{constant} & \mbox{if $\gamma > 0$;} \\
    0               & \mbox{otherwise.}
  \end{array}
  \right.
  \label{eq:Prior}
\end{equation}
If we assume that the uncertainties on the observational data are Gaussian, we can
obtain the likelihood from $\chi^2$ \citep[Equation~(3.65)]{sivia06}:
\begin{equation}
  \likelihood \propto \exp \left( -\frac{\chi^2}{2} \right).
  \label{eq:Likelihood}
\end{equation}
Note that $\chi^2$ is a function of the model parameters. Thus, from Bayes' Theorem
(Equation~(\ref{eq:Bayes})) we obtain
\begin{equation}
  \posterior \propto \exp \left( -\frac{\chi^2}{2} \right),
  \label{eq:Bayes2}
\end{equation}
provided $\gamma>0$.

The standard method for obtaining a 90\%\ confidence interval is to vary the model parameter in
question from its best-fit value until $\chi^2$ has increased by 2.706 from its minimum value,
\chisqmin\ \citep{lampton76,avni76}. Therefore, if we assume that the $\chi^2$ surface is parabolic
near the minimum,
\begin{equation}
  \chi^2(\gamma) = \chisqmin + 2.706 \left( \frac{\gamma - \gamma_0}{\gamma_{L,U} - \gamma_0} \right)^2,
  \label{eq:chisq}
\end{equation}
where $\gamma_0$ is the best-fit value of $\gamma$, $\gamma_L$ is the lower limit of the confidence
interval (used if $\gamma\le\gamma_0$) and $\gamma_U$ is the upper limit of the confidence
interval (used if $\gamma>\gamma_0$).

Substituting Equation~(\ref{eq:chisq}) into Equation~(\ref{eq:Bayes2}), we obtain
\begin{equation}
  \hspace{-5mm}
  \posterior = \left\{
  \begin{array}{ll}
    0 & \mbox{if $\gamma\le0$;} \\
    \mathcal{N} \exp \left[ -1.353 \left( \dfrac{\gamma - \gamma_0}{\gamma_L - \gamma_0} \right)^2 \right] & \mbox{if $0<\gamma\le\gamma_0$;} \\
    \mathcal{N} \exp \left[ -1.353 \left( \dfrac{\gamma - \gamma_0}{\gamma_U - \gamma_0} \right)^2 \right] & \mbox{if $\gamma>\gamma_0$,} \\
  \end{array}
  \right.
  \label{eq:Posterior}
\end{equation}
where
\begin{align}
  \mathcal{N} \equiv \Bigg\{ & \int_0^{\gamma_0}      \exp \left[ -1.353 \left( \dfrac{\gamma - \gamma_0}{\gamma_L - \gamma_0} \right)^2 \right] d\gamma +  \nonumber \\
                             & \int_{\gamma_0}^\infty \exp \left[ -1.353 \left( \dfrac{\gamma - \gamma_0}{\gamma_U - \gamma_0} \right)^2 \right] d\gamma \Bigg\}^{-1}
\end{align}
is the normalization. Note that Equation~(\ref{eq:Posterior}) is the posterior probability
distribution for $\gamma$ under the assumption that the exponential halo models are accurate
descriptions of the hot halo. We are not attempting to compare the probabilities of different types
of halo models, in order to determine which is the best description of the observational data.

\begin{figure}
  \centering
  \plotone{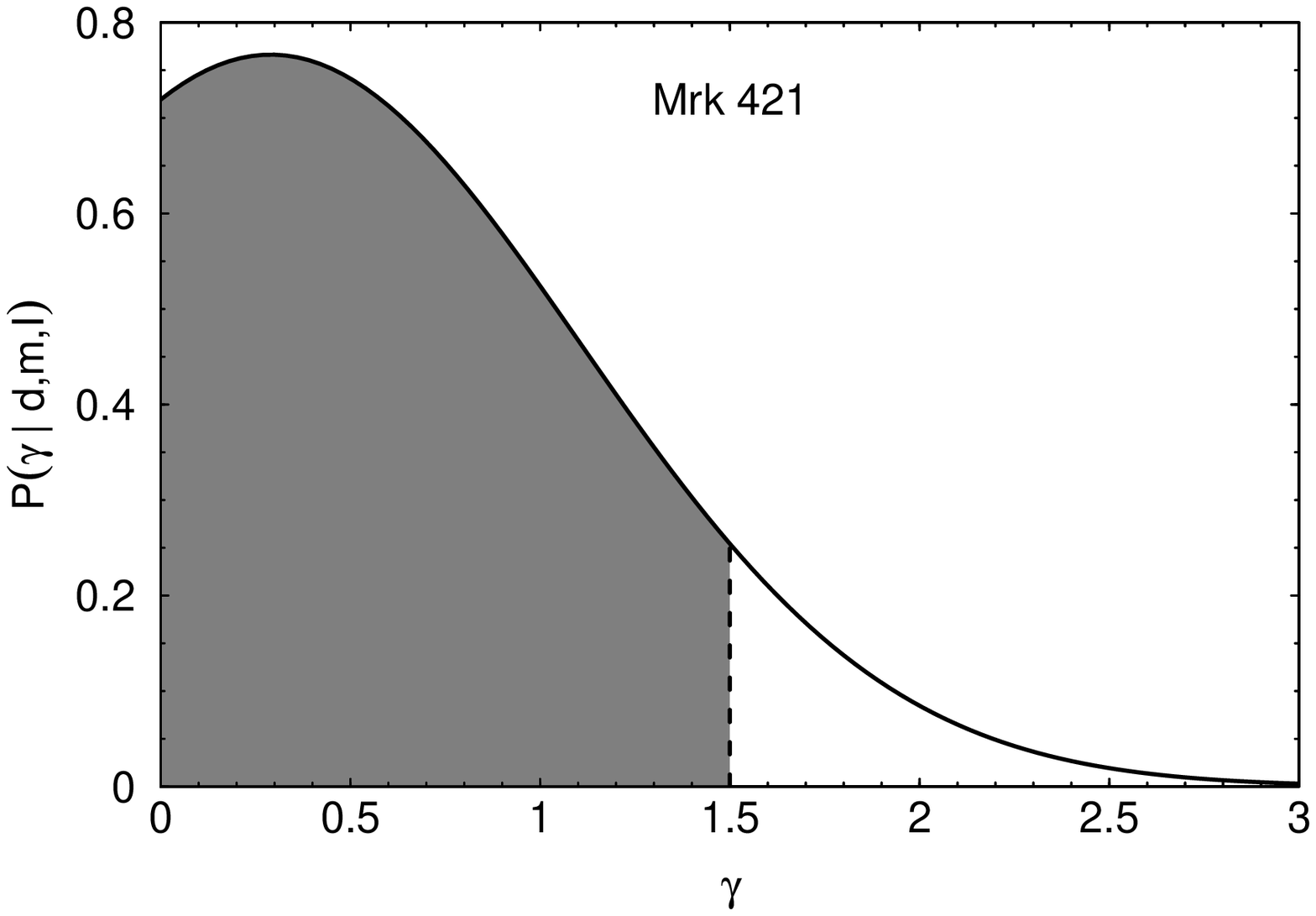} \\
  \plotone{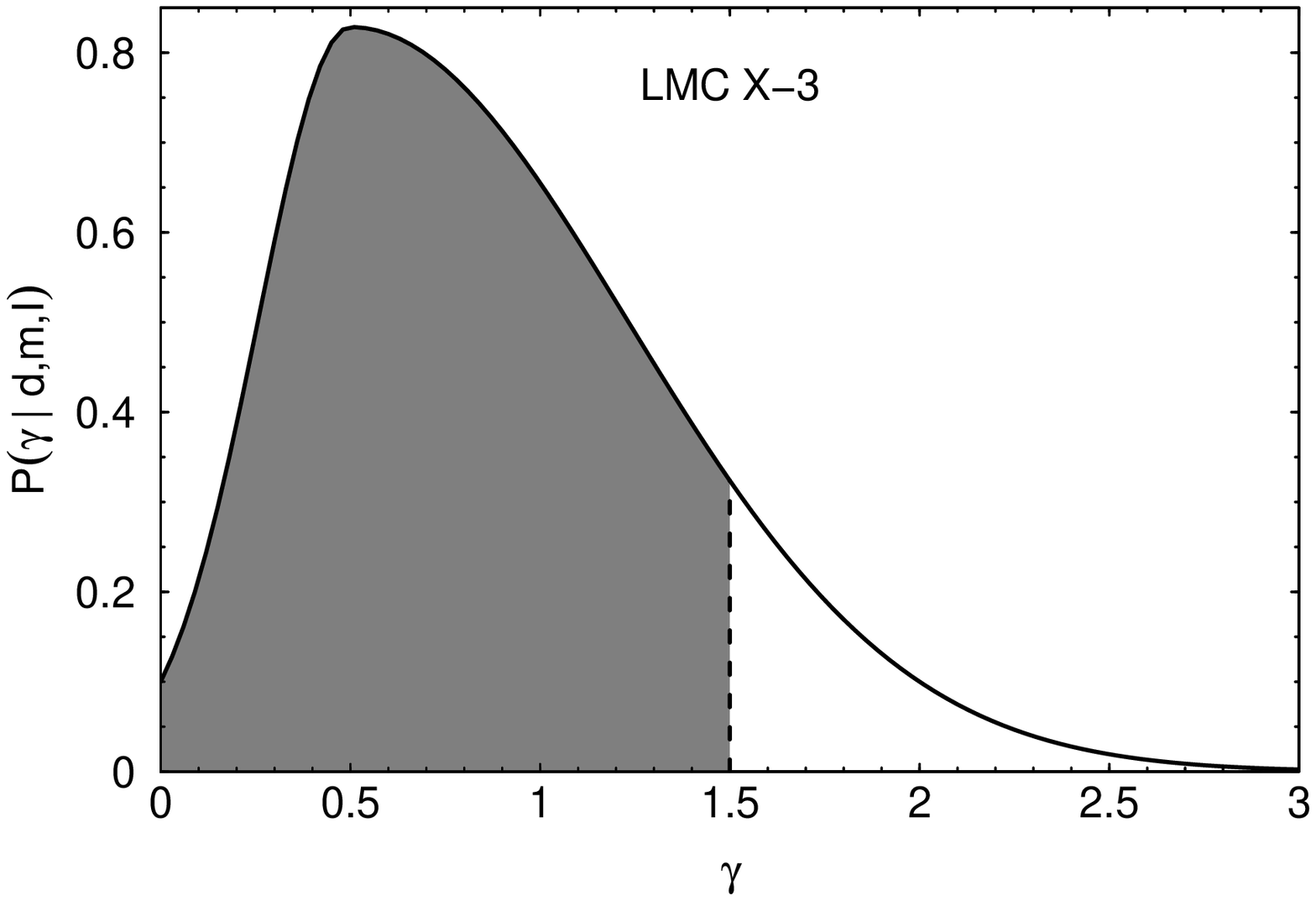} \\
  \plotone{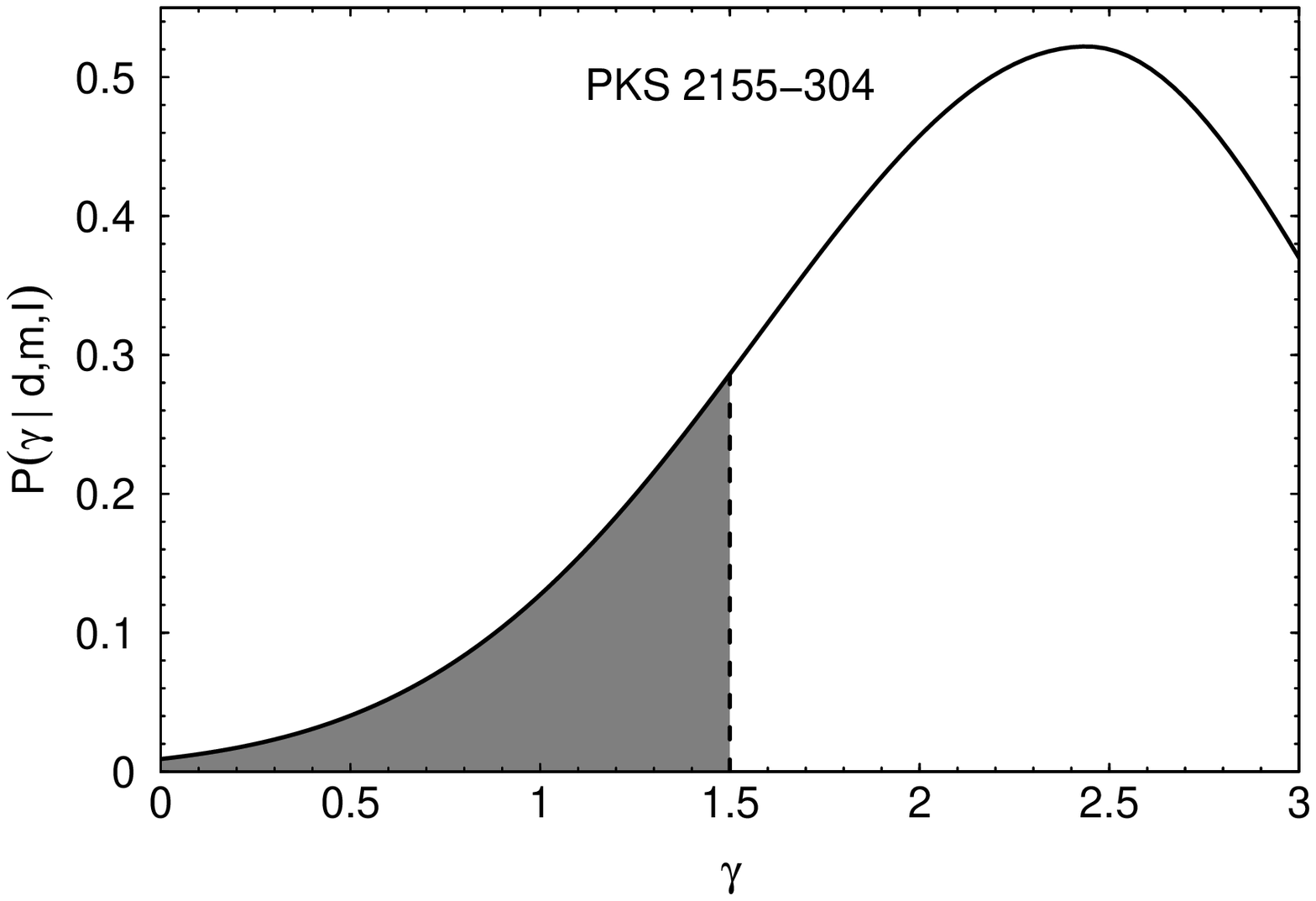}
  \caption{Posterior probability distributions for $\gamma$, calculated from the observations in Table~\ref{tab:Obs}
    using Equation~(\ref{eq:Posterior}). The shading indicates the region where $\gamma<3/2$, indicating that
    the hot halo is convectively unstable (Equation~(\ref{eq:Instability})).
    \label{fig:Posterior}}
\end{figure}

The posterior probability distributions for $\gamma$ derived from the observations in
Table~\ref{tab:Obs} are plotted in Figure~\ref{fig:Posterior}. Note that for the Mrk~421 sight line,
in the absence of any information on $\chi^2$ below $\gamma_0$, we assumed that the $\chi^2$ surface
is symmetrical about $\gamma_0$.

In each panel of Figure~\ref{fig:Posterior} we have shaded the region where $\gamma<3/2$ -- the
area of this region gives us the probability that the hot halo is convectively unstable, under the
assumption that the exponential halo models are accurate descriptions of the hot halo. These
probabilities are shown in the final column of Table~\ref{tab:Obs}. On the Mrk~421 and LMC~X-3 sight
lines, there is a slightly less than 90\%\ chance that the hot halo is convectively unstable. On the
PKS~2155$-$304 sight line, however, the probability of convective instability is much lower.

\section{EXTENDED HOT HALO MODELS}
\label{sec:Extended}

Typically, observational analyses that have concluded that the Milky Way's hot halo is extended have
assumed that the halo plasma is isothermal \citep{gupta12,gupta13,miller13}. While
\citet{gupta12,gupta13} assumed that the halo is of uniform density, in \citepossessive{miller13}
halo model the density decreases with Galactocentric distance (see Equation~(\ref{eq:BetaModel})),
in which case the halo plasma will be convectively stable.

\citet{fang13} examined several different models for the hot halo. They found that both an
exponential disk model (of the type discussed in the previous section) and a more extended
non-isothermal halo in hydrostatic equilibrium with the Galaxy's dark matter halo \citep{maller04}
could be made consistent with the existing X-ray emission and pulsar dispersion measure
data. \citet{fang13} argued that indirect evidence (the lack of gas in dwarf satellite galaxies and
the possible pressure-confinement of high-velocity clouds) favored the extended
\citeauthor{maller04}-like halo (which \citeauthor{fang13}\ refer to as the MB model).

The extended MB halo consists of adiabatic gas, with a polytropic index of 5/3. This means that the
gas is isentropic, placing it on the boundary between convective stability and instability. However,
in the absence of additional energy injection, radiative cooling would cause the entropy of
\citepossessive{fang13} MB model halo to change in such a way as to make the halo convectively
unstable, as we will now show.

\begin{figure}
  \centering
  \plotone{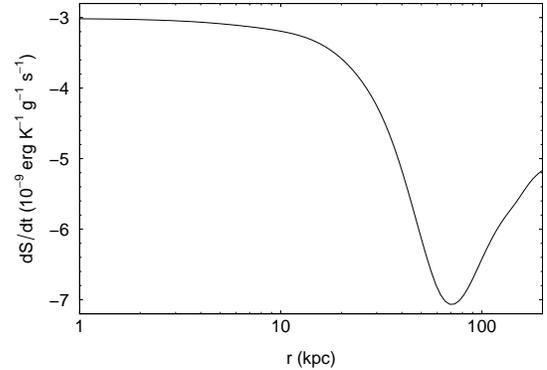}
  \caption{Rate of change of halo entropy due to radiative cooling as a function of Galactocentric
    distance, for \citepossessive{fang13} MB halo model (calculated using
    Equation~(\ref{eq:CoolingEntropy})).
    \label{fig:CoolingEntropy}}
\end{figure}

The change in the specific entropy, $d\mathcal{S}$, due to a radiative loss of energy $dQ$ is
$d\mathcal{S} = -dQ / T$. In time $dt$, the energy lost to radiative cooling is $n^2 \Lambda(T) dt$
per unit volume, or $(n/\bar{m}) \Lambda(T) dt$ per unit mass, where $\Lambda(T)$ is the cooling
function \citep[e.g.,][]{sutherland93}, and $\bar{m} \approx 1 \times 10^{-24}~\gram$ is the average
mass per particle. Hence,
\begin{equation}
  \frac{d\mathcal{S}}{dt} = - \frac{n \Lambda(T)}{\bar{m} T}.
  \label{eq:CoolingEntropy}
\end{equation}
Using the density and temperature profiles described by \citepossessive{fang13} Equations~(1) and
(2), normalized such that $n = 10^{-3.5}~\pcc$ and $T = 3.5 \times 10^6~\K$ at $r = 1~\kpc$
\citep[Figure~1]{fang13}, we plot $d\mathcal{S}/dt$ as a function of Galactocentric radius in
Figure~\ref{fig:CoolingEntropy}.

Figure~\ref{fig:CoolingEntropy} shows that radiative cooling decreases the entropy of the MB model
halo more rapidly at larger radii, out to $r \sim 70~\kpc$. Hence, while the model halo is assumed
to be isentropic, Figure~\ref{fig:CoolingEntropy} implies that this situation would not persist --
radiative cooling would result in the entropy decreasing with $r$, and the halo would become
convectively unstable. However, it should be noted that this change would occur slowly -- the cooling
time,
\begin{equation}
  t_\mathrm{cool} = \frac{3kT}{2n \Lambda(T)},
  \label{eq:CoolingTime}
\end{equation}
of the MB model plasma is $\sim$1--2~Gyr. In addition, radiative cooling would change both the
density and temperature structure of the halo -- a detailed simulation would be required to check if
the assumed density and temperature structure of the MB model halo would indeed ultimately give rise
to convection.

\section{DISCUSSION}
\label{sec:Discussion}

\subsection{The Observed Values of $\gamma$}

When considering the exponential halo model (Section~\ref{sec:Exponential}), we used published
values of $\gamma$ that were derived from X-ray data alone. \citet{yao09} also investigated
including \fuse\ \OVI\ absorption line data in their analysis of the LMC~X-3 sight line. They found
that it tightened the constraint on $\gamma$, from $0.5^{+1.2}_{-0.4}$ to $0.8^{+0.3}_{-0.4}$. This
appears to strengthen the conclusion that the halo is convectively unstable in this
direction. However, for the following reasons, we excluded \OVI\ from our analysis. In these
exponential models, the \OVI\ would reside several kiloparsecs above the disk.  This contradicts the
observation that $\sim$1/5 to $\sim$1/4 of the \OVI\ column density originates below $|z| = 1$~kpc
\citep{bowen08}. In addition, \citet{hagihara10} point out that, because \OVI-emitting plasma has a
high cooling rate, it may be difficult to maintain an \OVI-rich plasma high above the disk (although
a detailed calculation would be needed to determine the extent to which this is true). Therefore,
while including the \OVI\ absorption data can apparently tighten the constraint on $\gamma$, this
constraint may be unreliable.

The published values of $\gamma$ depend in part on the separation of the halo X-ray emission from
the other components of the soft X-ray background (SXRB), namely, the foreground emission from the
Local Bubble and/or from solar wind charge exchange, and the extragalactic background emission.  For
the PKS~2155$-$304 sight line, \citet{hagihara10} examined the effect of varying the normalization
of the foreground component of their SXRB emission model -- they tried both a normalization of zero,
and a normalization $\sim$75\%\ larger than their standard value. Decreasing (increasing) the
foreground normalization decreased (increased) the halo temperature inferred from the emission
spectrum, \Temis\ \citep[Table~4]{hagihara10}.

The lower value of \Temis\ resulting from a foreground normalization of zero is consistent with
\Tabs\ for this sight line, implying that the halo is isothermal on this sight line. Indeed, with a
foreground normalization of zero, the exponential halo model yields a relatively large value of
$\gamma$ (3.39, versus 2.44 for their standard foreground model). The temperature scale height being
a few times the density scale height implies that, from the point of view of the plasma relevant to
X-ray observations, the halo is close to being isothermal. Since the density decreases with height,
the halo is likely to be convectively stable in this case: with a foreground normalization of zero,
\citepossessive{hagihara10} results imply $P(\gamma<3/2) = 0.05$, compared with 0.15 for their
standard foreground normalization (Table~\ref{tab:Obs}). It should be noted, however, that a
foreground normalization of zero may be unrealistic -- models of solar wind charge exchange emission
suggest that $\ga$1~\lineunit\ of foreground \OVII\ emission is present in most observed SXRB
spectra \citep{koutroumpa07}, while \suzaku\ observations of the SXRB suggest a uniform foreground
\OVII\ intensity of $\sim$2~\lineunit\ (\citealt{yoshino09}; note that \citet{hagihara10} chose
their standard foreground normalization to yield an \OVII\ intensity of 2~\lineunit).

At the other extreme, increasing \citepossessive{hagihara10} foreground normalization led to a
greater discrepancy between \Temis\ and \Tabs. This led to a smaller best-fit value of $\gamma$
(1.47), and to an increased probability of convective instability: $P(\gamma<3/2) = 0.69$

\citet{yao09} also examined variants of their standard SXRB spectral model for the LMC~X-3 sight
line.  They varied the normalization of the foreground component and the spectrum of the
extragalactic component, and found that $\gamma$ was not strongly affected. However, these
variations on \citepossessive{yao09} analysis all included the \OVI\ absorption data. It is unclear
how sensitive the value of $\gamma$ derived for the LMC~X-3 sight line from X-ray data alone would
be to the other components of the SXRB. However, since $\gamma$ is better constrained for the
LMC~X-3 sight line than for the PKS~2155$-$304 sight line (compare rows~3 and 4 of
Table~\ref{tab:Obs}), varying the other components of the SXRB model is unlikely to have as large an
effect on $P(\gamma < 3/2)$ for the LMC~X-3 sight line as it did above for the PKS~2155$-$304 sight
line.

In summary, we acknowledge that the uncertainty in the foreground component of the SXRB emission
model introduces some uncertainty in our estimates of $P(\gamma<3/2)$ for the exponential halo
model. However, there is insufficient information in the published studies that we have used to
quantify this uncertainty.

The observed values of $\gamma$, and hence the probabilities of the halo being convectively
unstable that we have calculated for the exponential halo model, may be refined in the light of
future observations. In addition, future spectral analyses could use a Markov Chain Monte Carlo
method \citep[e.g.,][Section~15.8]{press07} to explore the model parameter space. Using such a
method would allow one to obtain the posterior distribution for $\gamma$ directly from the spectral
analysis, rather than having to infer it from the confidence intervals derived by varying
$\chi^2$. Furthermore, such an approach could allow one to marginalize over the normalization of the
foreground component of the SXRB emission model. This means that the posterior distribution for
$\gamma$ (and hence the probability of convective instability) would automatically take into account
the uncertainty in the foreground emission.

\subsection{Implications for the Hot Halo}

The observational results based on the exponential halo model imply that the probability that the
halo is convectively unstable in the direction of PKS~2155$-$304 is much lower than in the
directions of LMC~X-3 or Mrk~421 (Table~\ref{tab:Obs}). Since the halo is observed to be non-uniform
in emission \citep[e.g.,][]{yoshino09,henley10b,henley13a}, it is entirely possible for some regions
of the hot halo to be convectively unstable and for others to be stable. Note that PKS~2155$-$304 is
73\degr\ and 164\degr\ from LMC~X-3 and Mrk~421, respectively, and so the regions of the halo that
these sight lines are probing are well separated.

For a given halo model, we can using Equation~(\ref{eq:CoolingTime}) to calculate the cooling time
of the halo gas as a function of height, $z$, or Galactocentric radius, $r$.  We can then use $z /
t_\mathrm{cool}(z)$ or $r / t_\mathrm{cool}(r)$ to estimate the speed at which the halo fluid must
convect upward in order to offset the radiative losses. We can compare this with the adiabatic sound
speed,
\begin{equation}
  v_\mathrm{sound} = \sqrt{\frac{5 k T}{3 \bar{m}}},
  \label{eq:SoundSpeed}
\end{equation}
to see if the radiative energy losses can be offset by subsonic convective heating.

\begin{figure}
\centering
\plotone{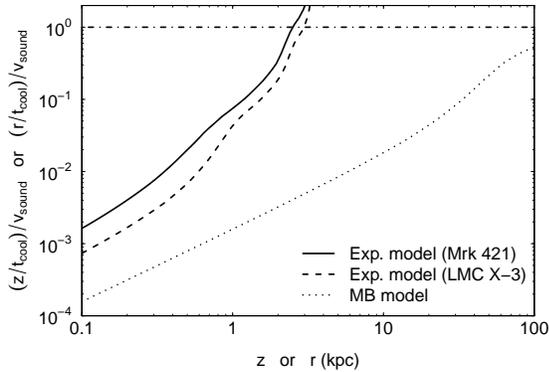}
\caption{The ratio of the estimated speed at which the halo fluid must convect upward in order to
  offset the radiative losses, $z / t_\mathrm{cool}$ or $r / t_\mathrm{cool}$, to the adiabatic
  sound speed, $v_\mathrm{sound}$. This ratio was calculated using Equations~(\ref{eq:CoolingTime})
  and (\ref{eq:SoundSpeed}).  The solid and dashed curves were calculated using the best-fit
  exponential halo models for the Mrk~421 and LMC~X-3 sight lines, respectively
  \citep{sakai12,yao09}, and are plotted as functions of height, $z$. The dotted curve was
  calculated using \citepossessive{fang13} MB model, and is plotted as a function of Galactocentric
  distance, $r$. The horizontal dot-dash line indicates $z / t_\mathrm{cool}$ or $r /
  t_\mathrm{cool} = v_\mathrm{sound}$.
  \label{fig:CoolingTime}}
\end{figure}

Figure~\ref{fig:CoolingTime} shows $(z / t_\mathrm{cool}) / v_\mathrm{sound}$ as a function of $z$,
calculated using the best-fit exponential halo models for the Mrk~421 (solid) and LMC~X-3 (dashed)
sight lines \citep{sakai12,yao09}. Up to $z \sim 2.5~\kpc$, $(z / t_\mathrm{cool}) /
v_\mathrm{sound} < 1$. Since these models imply that most of the X-ray-emitting and \OVII- and
\OVIII-bearing gas resides below $\sim$2.5~kpc (at which height $T \sim 6 \times 10^5~\K$;
\citealt{yao09,sakai12}), this result implies that this region could be kept hot by subsonic
convection replacing cooling gas with hotter gas from below.

Figure~\ref{fig:CoolingTime} also shows $(r / t_\mathrm{cool}) / v_\mathrm{sound}$ for the MB model
discussed in Section~\ref{sec:Extended} \citep{fang13}. For this model, $(r / t_\mathrm{cool}) /
v_\mathrm{sound} < 1$ out to a Galactocentric distance of at least 100~\kpc. The main reason $(r /
t_\mathrm{cool}) / v_\mathrm{sound}$ is small for this model is that the cooling time is long, as
noted in Section~\ref{sec:Extended}.  Hence, if radiative cooling perturbed the MB model, resulting
in a convective instability (Section~\ref{sec:Extended}), subsonic convection should be able to
maintain the temperature of the halo, in spite of radiative losses.

When considering the exponential models of the halo, we reiterate the point made in
Section~\ref{subsec:Probability} that the probabilities that the halo is convectively unstable
(Table~\ref{tab:Obs}) are conditional upon these exponential models being good descriptions of the
hot halo. We noted in Section~\ref{subsec:ModelSummary} that, in reality, the halo may consist of an
exponential-like distribution in the lower halo, plus a more extended, low-density halo suggested by
indirect evidence and expected from disk galaxy formation simulations. We do not attempt to come up
with a composite halo model here, but point out that the instability criterion derived in
Section~\ref{subsec:Instability} may be applied to the exponential portion of such a model.

Another issue that we do not attempt to address is the timescale on which convection would tend to
change the density and temperature distributions of the model halos we have examined.  Could a
dynamical equilibrium exist, or would convection tend to smooth out the temperature distribution on
a short timescale? If this timescale is short, it may argue against certain convectively unstable
models being accurate descriptions of the hot halo gas. Alternatively, one could require that a
model halo be convectively stable. In the case of the exponential halo model, this would involve
imposing a lower limit of 3/2 on $\gamma$ in the spectral analysis. More generally, it would involve
requiring that a model halo's entropy, $T n^{-2/3}$, increases with distance from the Galaxy. If, on
the other hand, convectively unstable halo models are good descriptions of the hot halo gas, in the
lower halo at least, this argues in favor of this gas being heated from the bottom, by supernova
activity in the disk.

\section{SUMMARY}
\label{sec:Summary}

We have examined the convective stability of the Milky Way's hot halo. Halo models in which the
density and temperature decrease exponentially with height \citep{yao07a,yao09,hagihara10,sakai12}
are convectively unstable if $\gamma<3/2$, where $\gamma$ is the ratio of the temperature and
density scale heights (Section~\ref{subsec:Instability}). Using the published best-fit values and
confidence intervals for $\gamma$, derived from joint analyses of X-ray emission and absorption line
data (Section~\ref{subsec:Observations}), we calculated the posterior probabilities for the hot halo
being convectively unstable, under the assumption that these exponential models are good
descriptions of the halo (Section~\ref{subsec:Probability}). We found that these probabilities are
just under 90\%\ in the directions of LMC~X-3 and Mrk~421 \citep{yao09,sakai12}, but only 15\%\ in
the direction of PKS2155$-$304 \citep{hagihara10}.  These results imply that, if the published
exponential models are good descriptions of the hot gas distribution (at least in the lower halo),
this gas is reasonably likely to be convectively unstable in two out of three directions, arguing in
favor of it being heated from the bottom by supernova activity in the disk.

We also examined model distributions in which the hot halo gas is more extended
(Section~\ref{sec:Extended}).  A variety of such models exists. \citet{miller13} assumed an
isothermal halo in which the gas density is described by a $\beta$-model in their analysis of
\OVII\ absorption line data. This model halo is convectively stable, as the temperature of the gas
is constant while its density decreases with distance from the Galaxy.  \citet{fang13}, meanwhile,
showed that a model in which the hot halo is assumed to be a non-isothermal gas in hydrostatic
equilibrium with the Galaxy's dark matter \citep{maller04} is consistent with the existing pulsar
dispersion measure and X-ray emission data. The gas in this model is isentropic, and would thus be
on the border between convective stability and instability if radiative cooling were
unimportant. However, we found that radiative cooling could perturb this model toward instability,
and hence that heating from disk supernovae via convection could play a role in maintaining such a
halo against radiative heat loss.

\acknowledgements

We thank the anonymous referee, whose comments helped significantly improve this paper.
Support for this work was provided by the National Aeronautics and Space Administration through
\chandra\ Award Number AR2-13017X issued by the \chandra\ X-ray Observatory Center, which is
operated by the Smithsonian Astrophysical Observatory for and on behalf of the National Aeronautics
Space Administration under contract NAS8-03060.

\bibliography{references}

\end{document}

%% file: newcom.tex
\newcommand{\OVI}{\ion{O}{6}}
\newcommand{\OVII}{\ion{O}{7}}
\newcommand{\OVIII}{\ion{O}{8}}

\newcommand{\Ne}{\ensuremath{n_{\mathrm{e}}}}

\newcommand{\mA}{\ensuremath{\mbox{m\AA}}}

\newcommand{\nm}{\ensuremath{\mbox{\nm}}}
\newcommand{\cm}{\ensuremath{\mbox{cm}}}

\newcommand{\kpc}{\ensuremath{\mbox{kpc}}}

\newcommand{\s}{\ensuremath{\mbox{s}}}

\newcommand{\gram}{\ensuremath{\mbox{g}}}

\newcommand{\sr}{\ensuremath{\mbox{sr}}}

\newcommand{\K}{\ensuremath{\mbox{K}}}

\newcommand{\ph}{\ensuremath{\mbox{photons}}}

\newcommand{\pcc}{\ensuremath{\cm^{-3}}}
\newcommand{\pcmsq}{\ensuremath{\cm^{-2}}}

\newcommand{\ps}{\ensuremath{\s^{-1}}}
\newcommand{\psr}{\ensuremath{\sr^{-1}}}

\newcommand{\lineunit}{\ph\ \pcmsq\ \ps\ \psr}

\newcommand{\chandra}{\textit{Chandra}}

\newcommand{\fuse}{\textit{FUSE}}

\newcommand{\suzaku}{\textit{Suzaku}}

\newcommand{\xmm}{\textit{XMM-Newton}}

\newcommand{\citepossessive}[1]{\citeauthor{#1}'s \citeyearpar{#1}}

\providecommand{\eqref}[1]{equation~(\ref{#1})}

%% file: ms.bbl
\begin{thebibliography}{39}
\expandafter\ifx\csname natexlab\endcsname\relax\def\natexlab#1{#1}\fi

\bibitem[{Avillez \& Mac~Low(2001)}]{avillez01}
Avillez, M.~A., \& Mac~Low, M.-M. 2001, ApJ, 551, L57

\bibitem[{Avni(1976)}]{avni76}
Avni, Y. 1976, ApJ, 210, 642

\bibitem[{Bowen {et~al.}(2008)Bowen, Jenkins, Tripp, Sembach, Savage, Moos,
  Oegerle, Friedman, Gry, Kruk, Murphy, Sankrit, Shull, Sonneborn, \&
  York}]{bowen08}
Bowen, D.~V., Jenkins, E.~B., Tripp, T.~M., {et~al.} 2008, ApJS, 176, 59

\bibitem[{Bregman(1980)}]{bregman80}
Bregman, J.~N. 1980, ApJ, 236, 577

\bibitem[{Crain {et~al.}(2010)Crain, McCarthy, Frenk, Theuns, \&
  Schaye}]{crain10}
Crain, R.~A., McCarthy, I.~G., Frenk, C.~S., Theuns, T., \& Schaye, J. 2010,
  MNRAS, 407, 1403

\bibitem[{Fang {et~al.}(2013)Fang, Bullock, \& Boylan-Kolchin}]{fang13}
Fang, T., Bullock, J., \& Boylan-Kolchin, M. 2013, ApJ, 762, 20

\bibitem[{Gupta {et~al.}(2009)Gupta, Galeazzi, Koutroumpa, Smith, \&
  Lallement}]{gupta09b}
Gupta, A., Galeazzi, M., Koutroumpa, D., Smith, R., \& Lallement, R. 2009, ApJ,
  707, 644

\bibitem[{Gupta {et~al.}(2013)Gupta, Mathur, Galeazzi, \& Krongold}]{gupta13}
Gupta, A., Mathur, S., Galeazzi, M., \& Krongold, Y. 2013, ApJ, submitted
  (arXiv:1307.6195)

\bibitem[{Gupta {et~al.}(2012)Gupta, Mathur, Krongold, Nicastro, \&
  Galeazzi}]{gupta12}
Gupta, A., Mathur, S., Krongold, Y., Nicastro, F., \& Galeazzi, M. 2012, ApJ,
  756, L8

\bibitem[{Hagihara {et~al.}(2010)Hagihara, Yao, Yamasaki, Mitsuda, Wang, Takei,
  Yoshino, \& McCammon}]{hagihara10}
Hagihara, T., Yao, Y., Yamasaki, N.~Y., {et~al.} 2010, PASJ, 62, 723

\bibitem[{Henley \& Shelton(2008)}]{henley08a}
Henley, D.~B., \& Shelton, R.~L. 2008, ApJ, 676, 335

\bibitem[{Henley \& Shelton(2013)}]{henley13a}
Henley, D.~B., \& Shelton, R.~L. 2013, ApJ, 773, 92

\bibitem[{Henley {et~al.}(2010)Henley, Shelton, Kwak, Joung, \&
  Mac~Low}]{henley10b}
Henley, D.~B., Shelton, R.~L., Kwak, K., Joung, M.~R., \& Mac~Low, M.-M. 2010,
  ApJ, 723, 935

\bibitem[{Hill {et~al.}(2012)Hill, Joung, Mac~Low, Benjamin, Haffner,
  Klingenberg, \& Waagan}]{hill12}
Hill, A.~S., Joung, M.~R., Mac~Low, M.-M., {et~al.} 2012, ApJ, 750, 104

\bibitem[{Joung \& Mac~Low(2006)}]{joung06}
Joung, M.~K.~R., \& Mac~Low, M.-M. 2006, ApJ, 653, 1266

\bibitem[{Koutroumpa {et~al.}(2007)Koutroumpa, Acero, Lallement, Ballet, \&
  Kharchenko}]{koutroumpa07}
Koutroumpa, D., Acero, F., Lallement, R., Ballet, J., \& Kharchenko, V. 2007,
  A\&A, 475, 901

\bibitem[{Koutroumpa {et~al.}(2011)Koutroumpa, Smith, Edgar, Kuntz, Plucinsky,
  \& Snowden}]{koutroumpa11}
Koutroumpa, D., Smith, R.~K., Edgar, R.~J., {et~al.} 2011, ApJ, 726, 91

\bibitem[{Kuntz \& Snowden(2000)}]{kuntz00}
Kuntz, K.~D., \& Snowden, S.~L. 2000, ApJ, 543, 195

\bibitem[{Lampton {et~al.}(1976)Lampton, Margon, \& Bowyer}]{lampton76}
Lampton, M., Margon, B., \& Bowyer, S. 1976, ApJ, 208, 177

\bibitem[{Mac~Low {et~al.}(1989)Mac~Low, McCray, \& Norman}]{maclow89}
Mac~Low, M.-M., McCray, R., \& Norman, M.~L. 1989, ApJ, 337, 141

\bibitem[{Maller \& Bullock(2004)}]{maller04}
Maller, A.~H., \& Bullock, J.~S. 2004, MNRAS, 355, 694

\bibitem[{McCammon {et~al.}(2002)McCammon, Almy, Apodaca, Bergmann~Tiest, Cui,
  Deiker, Galeazzi, Juda, Lesser, Mihara, Morgenthaler, Sanders, Zhang,
  Figueroa-Feliciano, Kelley, Moseley, Mushotzky, Porter, Stahle, \&
  Szymkowiak}]{mccammon02}
McCammon, D., Almy, R., Apodaca, E., {et~al.} 2002, ApJ, 576, 188

\bibitem[{Miller \& Bregman(2013)}]{miller13}
Miller, M.~J., \& Bregman, J.~N. 2013, ApJ, 770, 118

\bibitem[{Nicastro {et~al.}(2002)Nicastro, Zezas, Drake, Elvis, Fiore,
  Fruscione, Marengo, Mathur, \& Bianchi}]{nicastro02}
Nicastro, F., Zezas, A., Drake, J., {et~al.} 2002, ApJ, 573, 157

\bibitem[{Norman \& Ikeuchi(1989)}]{norman89}
Norman, C.~A., \& Ikeuchi, S. 1989, ApJ, 345, 372

\bibitem[{Press {et~al.}(2007)Press, Teukolsky, Vetterling, \&
  Flannery}]{press07}
Press, W.~H., Teukolsky, S.~A., Vetterling, W.~T., \& Flannery, B.~P. 2007,
  {Numerical Recipes -- The Art of Scientific Computing}, 3rd edn. (Cambridge:
  Cambridge University Press)

\bibitem[{Rasmussen {et~al.}(2003)Rasmussen, Kahn, \& Paerels}]{rasmussen03}
Rasmussen, A., Kahn, S.~M., \& Paerels, F. 2003, in The IGM/Galaxy Connection.
  The Distribution of Baryons at $z=0$, ed. J.~L. Rosenberg \& M.~E. Putman
  (Dordrecht: Kluwer), 109

\bibitem[{Rasmussen {et~al.}(2009)Rasmussen, Sommer-Larsen, Pedersen, Toft,
  Benson, Bower, \& Grove}]{rasmussen09}
Rasmussen, J., Sommer-Larsen, J., Pedersen, K., {et~al.} 2009, ApJ, 697, 79

\bibitem[{Sakai {et~al.}(2012)Sakai, Mitsuda, Yamasaki, Takei, Yao, Wang, \&
  McCammon}]{sakai12}
Sakai, K., Mitsuda, K., Yamasaki, N.~Y., {et~al.} 2012, in AIP Conf. Proc.
  1427, \textit{SUZAKU} 2011: Exploring the X-ray Universe: \textit{Suzaku} and
  Beyond, ed. R.~Petre, K.~Mitsuda, \& L.~Angelini (Melville: AIP), 342

\bibitem[{Shapiro \& Field(1976)}]{shapiro76}
Shapiro, P.~R., \& Field, G.~B. 1976, ApJ, 205, 762

\bibitem[{Shu(1992)}]{shu92}
Shu, F.~H. 1992, {The Physics of Astrophysics, Volume II: Gas Dynamics}
  (Sausalito, CA: University Science Books)

\bibitem[{Sivia \& Skilling(2006)}]{sivia06}
Sivia, D.~S., \& Skilling, J. 2006, {Data Analysis: A Bayesian Tutorial}, 2nd
  edn. (Oxford: Oxford University Press)

\bibitem[{Smith {et~al.}(2007)Smith, Bautz, Edgar, Fujimoto, Hamaguchi, Hughes,
  Ishida, Kelley, Kilbourne, Kuntz, McCammon, Miller, Mitsuda, Mukai,
  Plucinsky, Porter, Snowden, Takei, Terada, Tsuboi, \& Yamasaki}]{smith07a}
Smith, R.~K., Bautz, M.~W., Edgar, R.~J., {et~al.} 2007, PASJ, 59, S141

\bibitem[{Sutherland \& Dopita(1993)}]{sutherland93}
Sutherland, R.~S., \& Dopita, M.~A. 1993, ApJS, 88, 253

\bibitem[{Toft {et~al.}(2002)Toft, Rasmussen, Sommer-Larsen, \&
  Pedersen}]{toft02}
Toft, S., Rasmussen, J., Sommer-Larsen, J., \& Pedersen, K. 2002, MNRAS, 335,
  799

\bibitem[{Wang \& Yao(2012)}]{wang12}
Wang, Q.~D., \& Yao, Y. 2012, arXiv:1211.4834

\bibitem[{Yao \& Wang(2007)}]{yao07a}
Yao, Y., \& Wang, Q.~D. 2007, ApJ, 658, 1088

\bibitem[{Yao {et~al.}(2009)Yao, Wang, Hagihara, Mitsuda, McCammon, \&
  Yamasaki}]{yao09}
Yao, Y., Wang, Q.~D., Hagihara, T., {et~al.} 2009, ApJ, 690, 143

\bibitem[{Yoshino {et~al.}(2009)Yoshino, Mitsuda, Yamasaki, Takei, Hagihara,
  Masui, Bauer, McCammon, Fujimoto, Wang, \& Yao}]{yoshino09}
Yoshino, T., Mitsuda, K., Yamasaki, N.~Y., {et~al.} 2009, PASJ, 61, 805

\end{thebibliography}
